\documentclass[twocolumn,amsmath,amssymb,prl,showpacs]{revtex4}
\usepackage{graphicx}
\usepackage{times}

\begin{document}

\title{Network Robustness: Detecting Topological Quantum Phases}

\author{Chung-Pin Chou}

\affiliation{Beijing Computational Science Research Center, Beijing
100084, China}

\begin{abstract}
Can the topology of a network that consists of many particles
interacting with each other change in complexity when a phase
transition occurs? The answer to this question is particularly
interesting to understand the nature of phase transitions if the
distinct phases do not break any symmetry, such as topological phase
transitions. Here we present a novel theoretical framework
established by complex network analysis for demonstrating that
across a transition point of the topological superconductors, the
network space experiences a homogeneous-heterogeneous transition
invisible in real space. This transition is nothing but related to
the robustness of a network to random failures. We suggest that the
idea of the network robustness can be applied to characterizing
various phase transitions whether or not the symmetry is broken.
\end{abstract}


\maketitle


Phases of matter can be distinguished by using Landau's approach,
which characterizes phases in terms of underlying symmetries that
are spontaneously broken.
The information we need to understand phase transitions is usually
encoded in appropriate correlation functions, e.g. the correlation
length would diverge close to a quantum critical point.
Particularly, the low-lying excitations and the long-distance
behavior of the correlations near the critical phase are believed to
be well described by quantum field theory.
A major problem is, however, that in some cases it is unclear how to
extract important information from the correlation functions if
these phases do not break any symmetries, such as topological phase
transitions \cite{WenBook06,NayakRMP08,AliceaRPP12,BernevigBook13}.

\begin{figure}[b]
\center
\includegraphics[height=2.6in,width=2.8in]{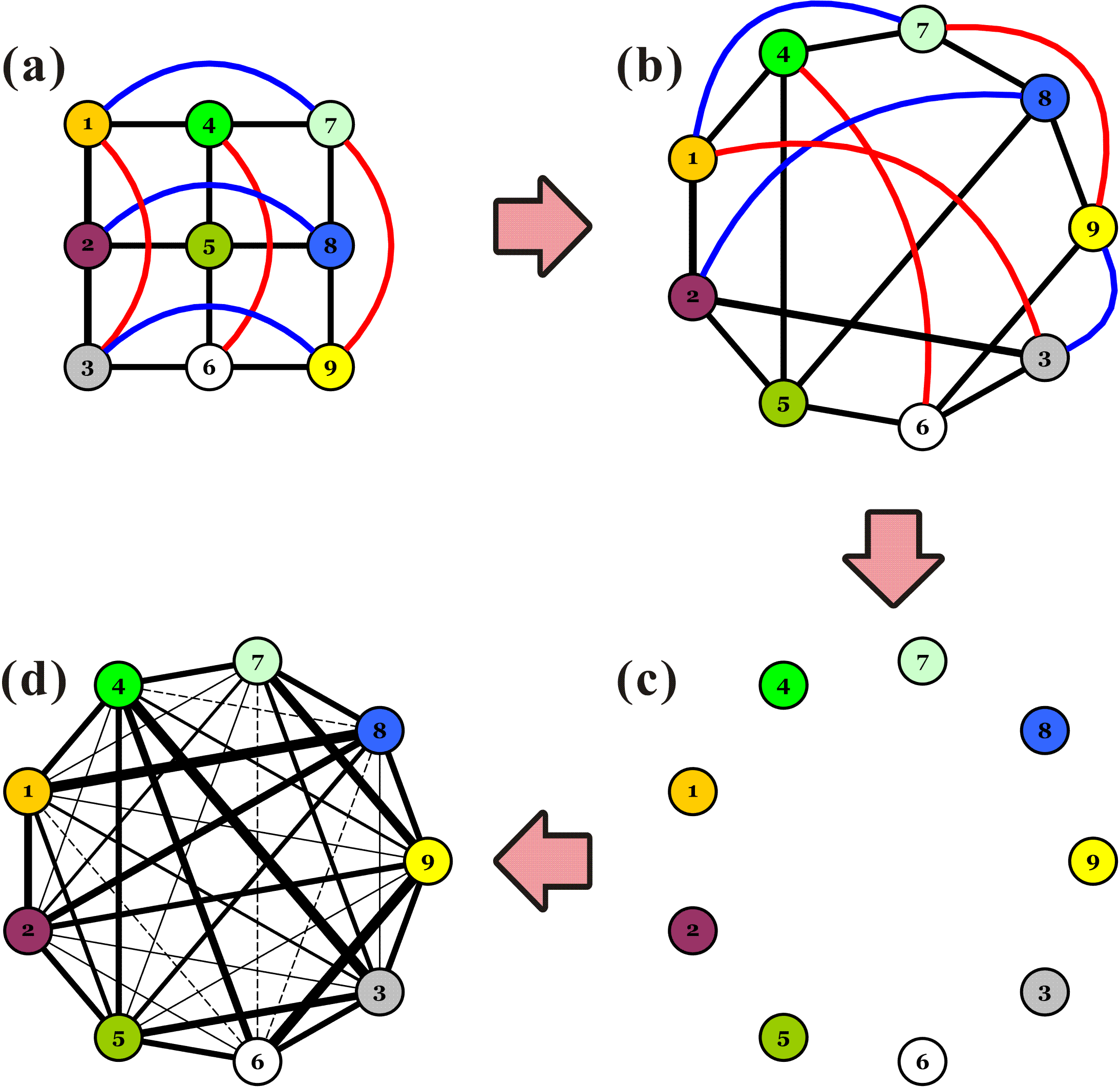}
\caption{An example of the construction of complex many-body
networks: (a) The system is a $3\times3$ square lattice with
periodic boundary conditions; (b) The system is replotted based on
the same topology as (a); (c) All links are removed from (b); (d)
All nodes are reconnected by using some correlations between
particles. The thickness of links denotes the magnitude of the
correlation.}\label{fig1}
\end{figure}

Complex network theory has become one of the most powerful
frameworks for understanding network structures of many real-world
systems
\cite{AlbertRMP02,DorogovtsevAIP02,NewmanSIAM03,BoccalettiPR06,DorogovtsevRMP08}.
According to graph theory, the elements of a system often are called
nodes and the relationships between them, which a weight is
associated with, are called links.
Decades ago, this unnoticed idea constructing a weighted network
from condensed matters had been proposed in quantum Hall systems
\cite{SenthilPRL99}.
In the language of network analysis, therefore, each network of $N$
nodes can be described by the $N\times N$ adjacency matrix
$\hat{A}$.
In what follows, we consider lattice sites as the nodes of the
weighted network of which each weighted link between nodes $i$ and
$j$ is expressed by the element of the adjacency matrix
$\hat{A}_{ij}$.
In Fig.\ref{fig1}(a), we start from a square-lattice example of size
$N=9$ in which the nodes form a simple regular network.
Following the procedure [Fig.\ref{fig1}(b) and (c)], we then
reconnect them by using some correlation functions as weights of
network links to generate a complete network with the link-weight
distribution (see Fig.\ref{fig1}(d)).
The links now carry the weights containing information about
important relationship between particles in many-body systems.

In this letter, focusing on the topological superconductors in one
(1D) and two (2D) dimensions \cite{KitaevPU01,ReadPRB00}, we explore
the possibilities of detecting the topological phase transitions by
using the novel network analysis.
Our analysis reveals that (i) a homogeneous-heterogeneous transition
occurs in network space from a topologically trivial phase to a
topologically non-trivial phase, which is accompanied by a hidden
symmetry breaking (namely, a reduction of the network robustness),
and (ii) the complex many-body network analysis can be applied to
other phase transitions without a prior knowledge of the system's
symmetry.

\textit{1D $p$-wave superconductor.}$-$The first model we consider
was introduced by Kitaev \cite{KitaevPU01}.
The Hamiltonian for $L$ spinless fermions in a chain with periodic
boundary conditions is
\begin{eqnarray}
H_{1D}=-\sum_{i}\left(\hat{c}_{i}^{\dag}\hat{c}_{i+1}+\hat{c}_{i}^{\dag}\hat{c}_{i+1}^{\dag}+H.c.\right)+\mu\hat{c}_{i}^{\dag}\hat{c}_{i},\label{KitaevR}
\end{eqnarray}
where $\mu$ is chemical potential.
The simplest superconducting (SC) model system shows the two-fold
ground-state degeneracy stemming from an unpaired Majorana fermion
at the end of the chain with open boundary conditions.
This model has two phases sharing the same physical symmetries: a
topologically trivial (strong pairing) phase for $\mu>\mu_{c}$($=2$)
and a topologically non-trivial (weak pairing) phase for
$\mu<\mu_{c}$.
The transition between them is the topological phase transition
identified by the presence or absence of unpaired Majorana fermions
localized at each end.

The Hamiltonian in momentum space is quadratic of fermionic
operators $\hat{c}_{\mathbf{k}}$, given by
\begin{eqnarray}
\sum_{\mathbf{k}}\left(
 \begin{array}{cc}
  \hat{c}_{\mathbf{k}}^{\dag} & \hat{c}_{-\mathbf{k}} \\
  \end{array}
  \right)\left(
         \begin{array}{cc}
         \epsilon_{\mathbf{k}} & -i\sin{\mathbf{k}} \\
         i\sin{\mathbf{k}} & -\epsilon_{\mathbf{k}} \\
         \end{array}
        \right)\left(
         \begin{array}{c}
         \hat{c}_{\mathbf{k}} \\
         \hat{c}_{-\mathbf{k}}^{\dag} \\
         \end{array}\right),\label{KitaevK}
\end{eqnarray}
where $\epsilon_{\mathbf{k}}=-\frac{\mu}{2}-\cos{\mathbf{k}}$.
By using the standard Bogoliubov transformation,
$\gamma_{\mathbf{k}}=\cos{(\theta_{\mathbf{k}}/2)}\hat{c}_{\mathbf{k}}-i\sin{(\theta_{\mathbf{k}}/2)}\hat{c}_{-\mathbf{k}}^{\dag}$
where
$\tan{\theta_{\mathbf{k}}}=\sin{\mathbf{k}}/\epsilon_{\mathbf{k}}$,
the Hamiltonian can be diagonalized.
The excitation spectrum of the form,
$E_{\mathbf{k}}=\sqrt{\left(2\epsilon_{\mathbf{k}}\right)^{2}+\sin^{2}{\mathbf{k}}}$,
remains fully gapped except at the critical point $\mu_{c}$.
The SC ground state is the state annihilated by all
$\gamma_{\mathbf{k}}$:
\begin{eqnarray}
|\Psi_{GS}\rangle=e^{\frac{1}{2}\sum_{i,j}G_{ij}\hat{c}_{i}^{\dag}\hat{c}_{j}^{\dag}}|0\rangle,\label{KitaevGS}
\end{eqnarray}
where $G_{ij}$ represents the pairing amplitude given by
\cite{ChungPRB00}
\begin{eqnarray}
G_{ij}=\frac{1}{L}\sum_{\mathbf{k}}\tan{(\theta_{\mathbf{k}}/2)}e^{i\mathbf{k}\cdot(\mathbf{r}_{i}-\mathbf{r}_{j})}.\label{KitaevGij}
\end{eqnarray}
A possible choice of the adjacency matrix of the 1D superconductor
is the normalized pairing amplitude in which the non-local property
between spinless fermions is concealed.
The adjacency matrix can serve as an intuitive definition for the
network of spinless fermions with $p$-wave Cooper pairing:
\begin{eqnarray}
\hat{A}_{ij}=\frac{|G_{ij}|}{\max{G_{ij}}}.\label{AijGij}
\end{eqnarray}
The node $i$ or $j$ stands for a given lattice site.
The weights of links contain information about the pairing strength
between spinless fermions.

\begin{figure}[t]
\center
\includegraphics[height=2.5in,width=3.2in]{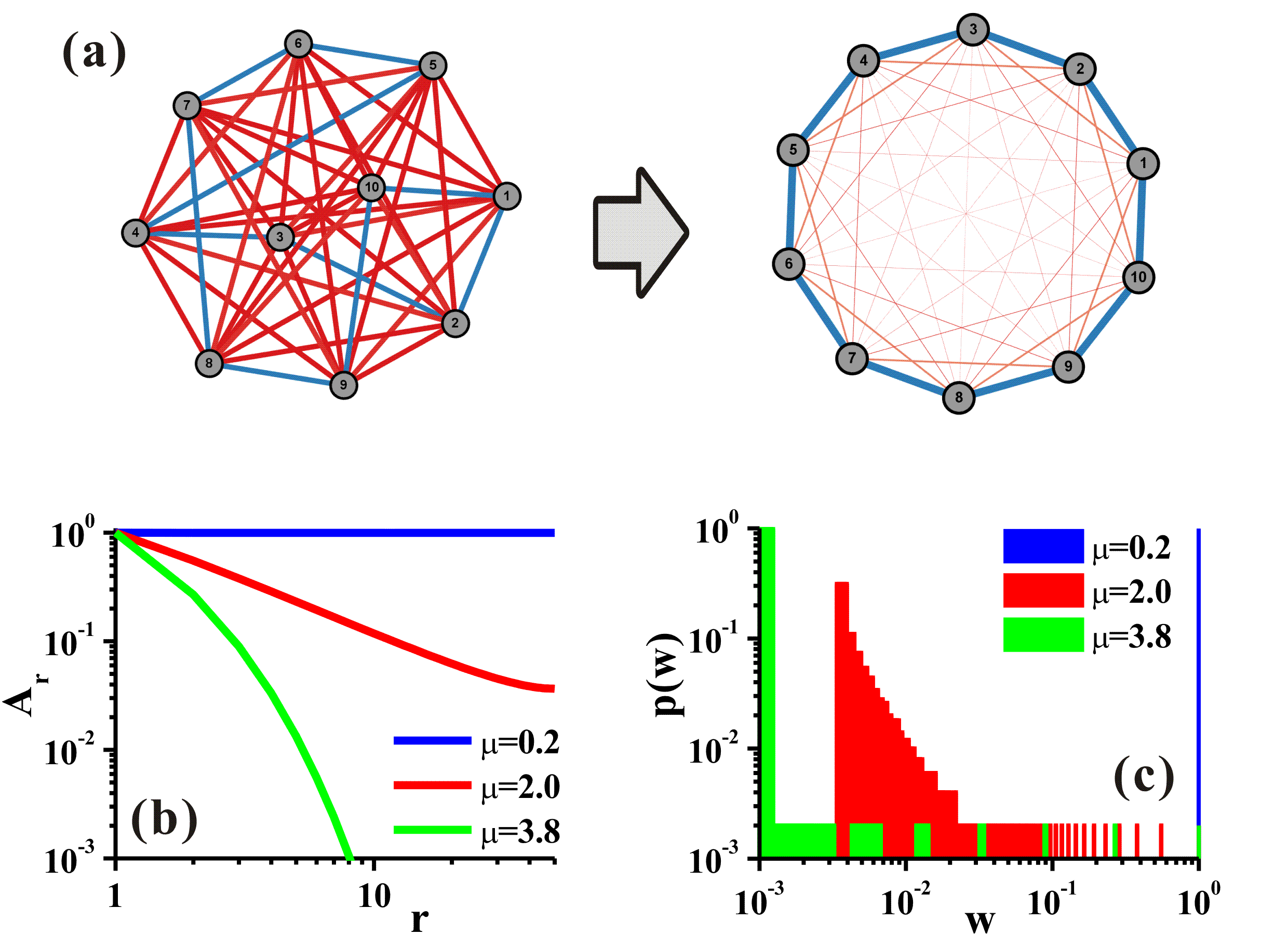}
\caption{(a) The evolution of network topologies of the 1D $p$-wave
superconductor from chemical potential $\mu=0.2$ to $4.2$. The chain
length $L=10$. The thickness and color of links represent their
weights. Color scale: Blue (Red) indicates the largest (smallest)
weights. (b) Pairing amplitude $A_{r}$ as a function of distance $r$
for different $\mu$. (c) The probability distribution $p(w)$ of the
weights $w$ of network links for different $\mu$. The bin size is
chosen for clear demonstrations. The chain length
$L=1000$.}\label{fig2}
\end{figure}

In Fig.\ref{fig2}(a), one can see the change of network topologies
for different $\mu$ in a short chain of $L=10$.
Below $\mu_{c}$, the topologically non-trivial phase displays
irregular patterns of the complete network, where each node is
connected to all other nodes with different link weights.
As increasing $\mu$ above $\mu_{c}$, the topologically trivial phase
demonstrates a ring structure comprised of the nodes with the
largest link weight in the network pond.
There are only few links with the strongest weight that is called
"highways" of the network.
The obvious change of topologies of the network across the critical
point is intimately related to the critical behavior observed in
real space.

We now recall the pairing amplitude in real space shown in
Eq.(\ref{KitaevGij}).
Consider translational invariance, Figure \ref{fig2}(b) shows how
the normalized pairing amplitude
$A_{r\equiv|\mathbf{r}_{i}-\mathbf{r}_{j}|}$($=\hat{A}_{ij}$)
changes as the topological phase transition occurs.
For $\mu<\mu_{c}$, the weak pairing phase indicates that the size of
the Cooper pair is infinite, leading to $A_{r}\sim const$.
At the critical point, the critical phase has power-law correlations
at large distances.
Above the critical point, {\it i.e.} $\mu>\mu_{c}$, the strong
pairing phase instead shows that the pairing amplitude is
exponentially decaying with distances: $A_{r}\sim e^{-r/\xi}$.
The Cooper pairs form molecules from two fermions bound in real
space over a length scale $\xi$.
The exponentially decaying pairing amplitude in real space results
in a ring structure in network space.
Similar physics would also appear in the well-known phenomena of
BEC-BCS crossover in $s$-wave superconductors \cite{ChenPR05}.

To further analyze the network structure, we examine how the
probability distribution $p(w)$ of the weights $w$ of network links
evolves over contiguous topological phases.
In Fig.\ref{fig2}(c), the weights distribute like a delta function
for $\mu<\mu_{c}$.
Namely, the weights homogeneously distribute in network space.
As further increasing $\mu$, the distributions begin to lose weight
but still remain nearly homogeneous.
It is noteworthy that the distribution at the critical point
possesses a decaying function with a heavy tail.
Hence the weight distribution of network links becomes more
heterogeneous.
In the strong pairing regime ($\mu>\mu_{c}$), the distribution moves
to almost zero weight and recovers a sharp peak at $w\sim0$.
The tail of the distribution in the strong pairing regime thus looks
much more heterogeneous than the weak pairing regime.

This observation reminds us of a well-known fact in real-world
networks that a network with the heterogeneous weight distribution
of links is robust to random failures \cite{AlbertNat00,WangCM05}.
The robustness of the network originating from its heterogeneity
seems to indicate a hidden symmetry in network space.
More precisely, the hidden symmetry describes a phenomenon that the
network function and structure remain unchanged or invariant under
random removal of its links.
Thus, there exists a homogeneous-heterogeneous network transition
hidden in the topological phase transition.
It may allow us to define a topological order parameter in network
space for identifying the phase transition without any local order
parameter.


\textit{2D $p$+i$p$ superconductor.}$-$Consider now a 2D
time-reversal symmetry breaking superconductor, the $p$+i$p$
superconductor for $N$ spinless fermions:
\begin{eqnarray}
H_{2D}=\sum_{\mathbf{k}}\varepsilon_{\mathbf{k}}\hat{c}_{\mathbf{k}}^{\dag}\hat{c}_{\mathbf{k}}+\left(\Delta_{\mathbf{k}}\hat{c}_{\mathbf{k}}^{\dag}\hat{c}_{-\mathbf{k}}^{\dag}+H.c.\right),\label{pipH}
\end{eqnarray}
where the single-particle dispersion
$\varepsilon_{\mathbf{k}}=-2(\cos k_{x}+\cos k_{y})-\mu$ and the gap
function $\Delta_{\mathbf{k}}=\sin k_{x}+i\sin k_{y}$.
For the spinless fermions, the gap function has odd parity symmetry,
$\Delta_{-\mathbf{k}}=-\Delta_{\mathbf{k}}$.
One can see that the excitation spectrum has gapless nodes at
time-reversal invariant momenta: $(0,0)$, $(0,\pi)$, $(\pi,0)$,
$(\pi,\pi)$.

\begin{figure}[t]
\center
\includegraphics[height=2.5in,width=3.2in]{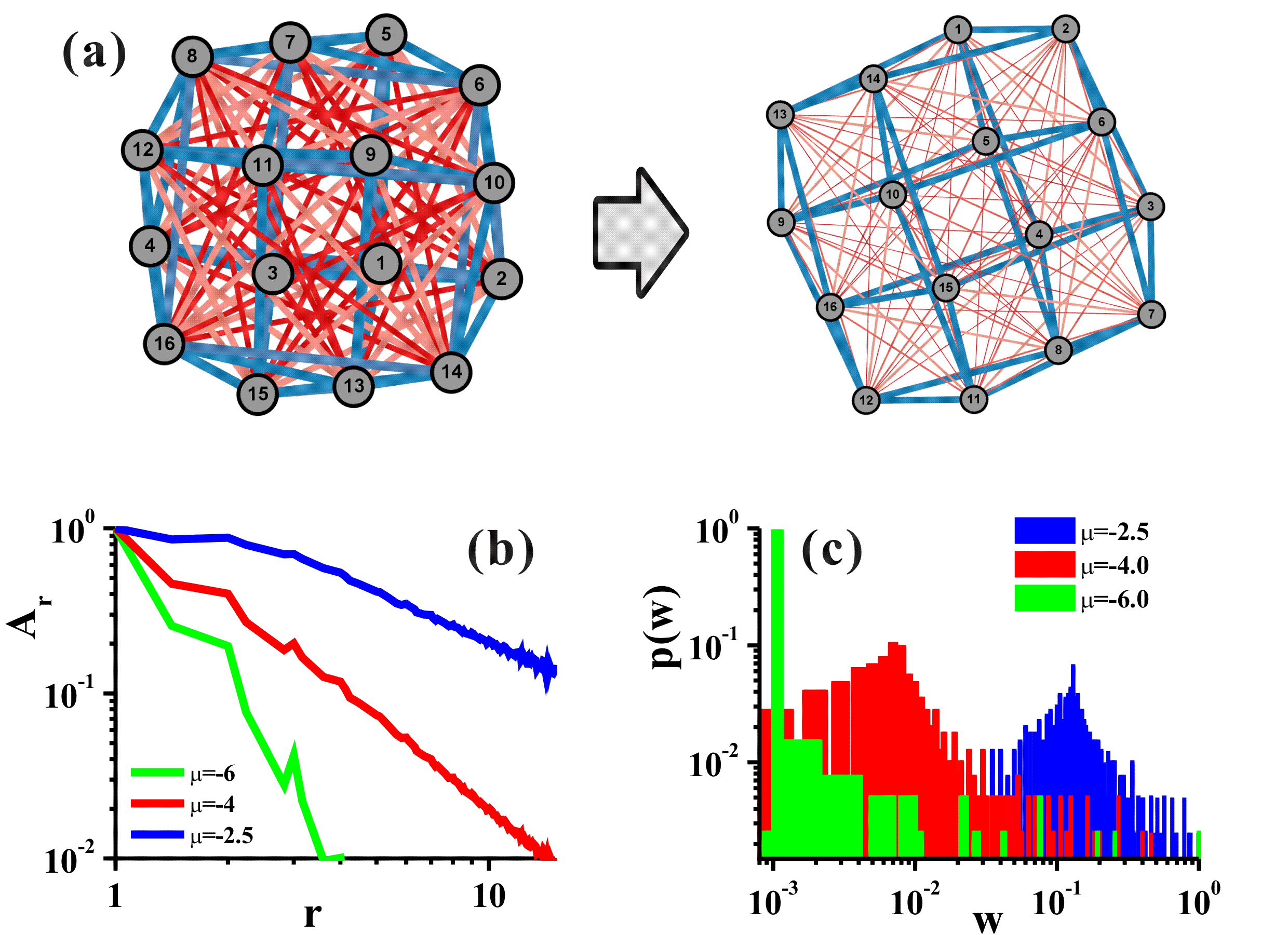}
\caption{(a) The evolution of network topologies of the 2D $p$+i$p$
superconductor from chemical potential $\mu=-2.5$ to $-6$. The
lattice size $N=16$. (b) Pairing amplitude $A_{r}$ as a function of
distance $r$ for different $\mu$. (c) The probability distribution
$p(w)$ of the weights $w$ of network links for different $\mu$. The
lattice size $N=1600$.}\label{fig3}
\end{figure}

Following the same analysis as the 1D example, the SC ground-state
wave function has the same form as Eq.(\ref{KitaevGS}).
However, the pairing amplitude $G_{ij}$ is now given by
\begin{eqnarray}
G_{ij}=\frac{1}{N}\sum_{\mathbf{k}}\frac{v_{\mathbf{k}}}{u_{\mathbf{k}}}e^{i\mathbf{k}\cdot(\mathbf{r}_{i}-\mathbf{r}_{j})},\label{pipGS}
\end{eqnarray}
where $v_{\mathbf{k}}$ and $u_{\mathbf{k}}$ are BCS coherence
factors (More details can be found in Ref.~\onlinecite{ReadPRB00}).
Note that the system preserves the particle-hole symmetry so that
only $\mu\leq0$ will be considered later.
This model also has two topological distinct phases: a topologically
trivial phase for $\mu<\mu_{c}$($=-4$) and a topologically
non-trivial phase for $\mu>\mu_{c}$.
Other than the 1D case, however, there is the other transition point
at $\mu'_{c}=0$ due to the bulk gap closure at $(\pi,0)$ and
$(0,\pi)$.

We now investigate how the network analysis performs in the face of
the 2D topological phase transition.
The definition for the adjacency matrix $\hat{A}_{ij}$ in the 2D
superconductor is still the same as Eq.(\ref{AijGij}).
The complex topologies of the weighted network for the topological
trivial and non-trivial phases are shown in Fig.\ref{fig3}(a) in a
square lattice of $N=16$.
The topologically non-trivial phase ($\mu>\mu_{c}$) gives rise to a
weighted complete network with the link-weight distribution.
It would be just a trivial complete network if the network were
unweighted.
For $\mu<\mu_{c}$, each node only has four highways to its
neighbors, hence the network topology is equivalent to a torus which
corresponds to a square lattice with periodic boundary conditions.

As in the 1D example the difference between the strong pairing phase
and the weak pairing phase can be distinguished by examining the
pairing amplitude.
In Figure \ref{fig3}(b), with $\mu>\mu_{c}$ a weakly paired
condensate forms from Cooper pairs loosely bound in real space,
which gives $A_{r}\sim r^{-1}$ at large distance.
As for $\mu<\mu_{c}$, the pairing amplitude falls exponentially for
large $r$, $A_{r}\sim e^{-r/\xi}$, because the pairs of the strong
pairing phase are tightly bound in real space.
Thus, the exponentially decaying pairing amplitude gives rise to the
torus structure in network space.

Let us turn to discussing the probability distribution of the
weights of network links.
For $\mu>\mu_{c}$, Fig.\ref{fig3}(c) shows that the distribution
still shows a bell-like function due to power-law decaying pairing
amplitude.
Most links centering around a moderate weight behave homogeneous in
network space.
Similar to the 1D case, the heterogeneity of the distribution
appears as further approaching $\mu_{c}$.
For $\mu<\mu_{c}$, a majority of links lose their weights, hence,
the distribution becomes much more heterogeneous and exhibits a long
tail.
The 2D superconductor shows much broader weight distribution of
network links than the 1D case.
Even so, a homogeneous-heterogeneous transition is still observed in
the network space.


\textit{Network measures.}$-$So far we have not introduced the
topological order parameter for the topological superconductors in
1D and 2D yet.
We turn our attention to two network measures that could be used in
these topological quantum systems.
One is the so-called small-world phenomena \cite{WattsBook99}.
Many of real-world networks have the property of relatively short
average path length defined by a shortest route running along the
links of a network.
A small-world network including not only strong clustering but also
short path length has also been introduced to describe real-world
networks \cite{WattsNat98}.

Instead of the weighted clustering coefficient $C$
\cite{OnnelaPRE05} and the average path length $D$
\cite{NewmanBook10} commonly used in network analysis (see more
details in the Supplemental Material \cite{suppl}), a measure of the
small-world property called "small-worldness" has been recently
proposed \cite{HumphriesPLos08,CPCArXiv13}.
It is defined as
\begin{eqnarray}
S\equiv\frac{C}{D},\label{smallworld}
\end{eqnarray}
which is based on the maximal tradeoff between high clustering and
short path length.
A network with larger $S$ has a higher small-world level.
The small-worldness seems to be appropriate to describing the
universal critical properties because it can extract information
about both locality (weighted clustering coefficient) and
non-locality (average path length) from network space.

\begin{figure}[t]
\center
\includegraphics[height=2.3in,width=3.5in]{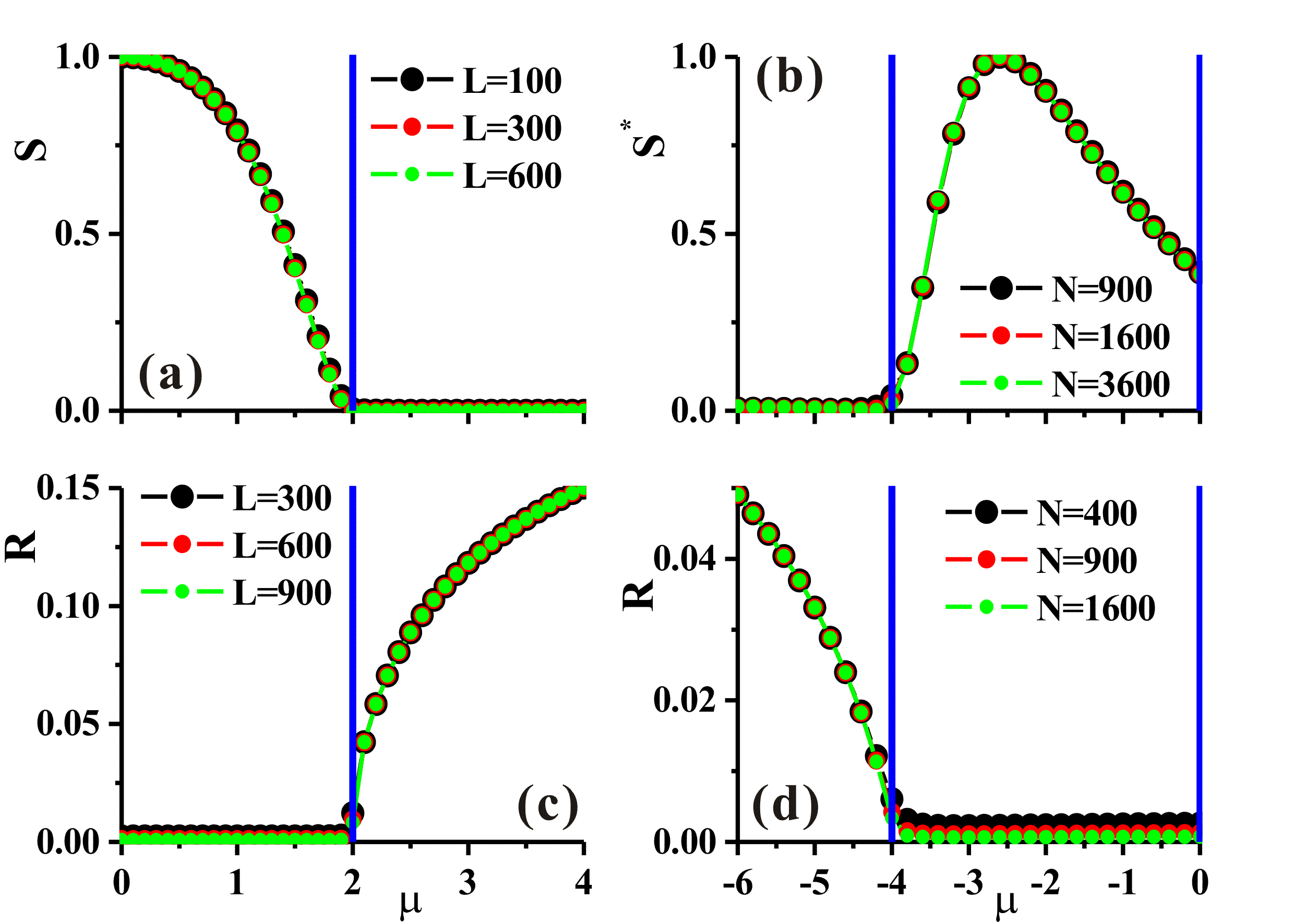}
\caption{(a) Small-worldness $S$ and (c) network robustness $R$ of
the 1D $p$-wave superconductor vs chemical potential $\mu$ for
different chain length $L$. (b) Normalized small-worldness $S^{*}$
and (d) network robustness $R$ of the 2D $p$+i$p$ superconductor for
different lattice size $N$ as a function of $\mu$. The blue lines
indicate the critical points.}\label{fig4}
\end{figure}

In Fig.\ref{fig4}(a), we illustrate the critical behavior of the
small-worldness in the 1D $p$-wave superconductor.
One can see that the small-worldness drops to zero when the 1D
superconductor comes from the topologically non-trivial phase to the
topologically trivial phase.
This coincidence convinces us that the network topology enables the
small-worldness, akin to an order parameter in the theory of
conventional phase transitions, to expose the change of nontrivial
topology inherent in the weak pairing regime \cite{CPCArXiv13}.
For the 2D $p$+i$p$ superconductor, however, the small-worldness
displays the notorious finite size effect for the topologically
non-trivial phase as a result of its much broader weight
distribution.
In order to overcome this hassle, we define the normalized
small-worldness as $S^{*}=S/\max S(\mu)$.
In Fig.\ref{fig4}(b), near the critical point $\mu_{c}$ the
normalized small-worldness vanishes in the strong pairing regime as
well.
Surprisingly, the other critical point $\mu'_{c}$ seems to be also
characterized by a decline in the normalized small-worldness.

There is an alternative way to understand the disappearance of the
small-worldness in the strong pairing regime.
As approaching the strong pairing phase, most links start to lose
weights and the weight distribution shows more heterogeneous, thus
leading to the reduction of the small-worldness.
The phenomenon that most links are like slow traffic lanes results
in vanishing small-worldness.
In other words, the regular network of the strong pairing phase has
a very small weighted clustering coefficient and much longer average
path length as the system size goes to infinity.
The same reasoning from the weight distribution of network links can
be also applied to other many-body systems with/without local order
parameters.
Hence, this result strongly suggests that the small-worldness can be
considered as a topological order parameter in the network space.

We have to mention a point now in passing.
In the network representation, there exists a hidden symmetry
corresponding to a heterogeneous network with a long-tail weight
distribution.
The homogeneous-heterogeneous transition of the network topology
observed is intimately related to the hidden symmetry breaking.
The hidden symmetry is nothing but the robustness of a complex
network.
The heterogeneity of network links implies that a weighted network
is more robust against random failures \cite{WangCM05}, accompanied
by higher hidden symmetry in network space.
Conversely, the homogeneity of network links means that a weighted
network becomes fragile to random failures, and thus breaks the
hidden symmetry.
The other network measure we have to take is just to quantify the
hidden symmetry.

The network structure and function strongly rely on its structural
robustness, i.e. the ability of a network to maintain the
connectivity when a fraction of nodes or links are randomly removed.
A variety of network measures have been proposed to detect the
structural robustness
\cite{HararyNet83,KrisCMA87,EsfIPL88,BauerCC90}.
Recently the concept of natural connectivity derived from the graph
spectrum has been introduced to measure the structural robustness
\cite{BarahonaCPL10,BarahonaChao12,EstradaPR12} (also see the
details in the Supplemental Material \cite{suppl}).
We can further extend the concept of natural connectivity to the
weighted network.
The natural connectivity in a weighted network represent the
"strength" of loops of all lengths instead of the number of loops.
Thus we call the natural connectivity as the network robustness $R$,
that can be given by
\begin{eqnarray}
R=\ln\left(\frac{1}{n}\sum_{i=1}^{n}e^{\bar{\lambda}_{i}}\right),\label{robustR}
\end{eqnarray}
where $\bar{\lambda}_{i}$($\equiv\lambda_{i}/\max\lambda_{i}$)
stands for the normalized eigenvalue of the adjacency matrix
$\hat{A}$ to avoid the enhancement of the network robustness as
increasing the number of nodes $n$($=L$ in 1D and $N$ in 2D).

In Fig.\ref{fig4}(c) and (d), we analyze the
homogeneous-heterogeneous transition of network topology by plotting
the network robustness $R$ for the topological superconductors.
As we expect, in both 1D and 2D cases the strong pairing phase
always exhibits more robust network structure than the weak pairing
phase, owing to its more heterogeneous weight distribution.
For the homogeneous-heterogeneous transition in network space, the
hidden symmetry breaking at the critical point indicates that the
network loses its robustness ($R=0$, namely, the hidden symmetry is
broken), further leading to an appearing order parameter: the
small-worldness ($S\neq0$).
This is a clear picture that Landau's symmetry breaking theory works
well even in network space.
More interestingly, the symmetry-breaking idea is successfully
applied to identifying the topological phase transitions in the
topological superconductors.
It is worthy to be mentioned that the concept behind the hidden
symmetry breaking can also provides significant information to
comprehend traditional phase transitions with Landau's symmetry
breaking in condensed matter systems.


\textit{Conclusions.}$-$By using complex network analysis we have
addressed how to read useful information from the pairing amplitude
to characterize the topological phases in 1D and 2D topological
superconductors.
We have illustrated that a network measure, small-worldness, plays a
significant role as a topological order parameter in network space,
relied on Landau symmetry-breaking picture.
The evolution of the weight distribution of network links across the
critical point is responsible for the change of the small-worldness,
which is analogous to the change of the speed limit on a road
network from the highway to the slow traffic lane.
The phenomenon that the structure of the weighted network varies
from heterogeneity to homogeneity implies a hidden symmetry
broken$-$or, to put it another way, the disappearance of the network
robustness to random failures.
The hidden symmetry breaking has been successfully described by
another network measure, network robustness.
The robustness of a complex network is able to uncover a wealth of
topological information underneath the pairing amplitude, and
further comprehend the mechanism of the phase transitions without
local order parameters.
We thus suggest that complex network analysis can be a valuable tool
to investigate quantum or classical phase transitions in condensed
matters.


We would like to thank Ting-Kuo Lee and Ming-Chung Chang for helpful
discussion and comments. Our special thanks go to Xiao-Sen Yang for
fruitful collaborations. This work is supported by CAEP and MST.

\end{document}